\documentclass[conference, 10pt]{IEEEtran}
\IEEEoverridecommandlockouts
\usepackage{cite}
\usepackage{amsmath,amssymb,amsfonts}
\usepackage{algorithmic}
\usepackage{graphicx}
\usepackage{textcomp}
\usepackage{xcolor}
\usepackage{multirow}
\usepackage{tabularx}

{
\usepackage{hyperref}

\makeatletter
\let\save@mathaccent\mathaccent
\newcommand*\if@single[3]{%
  \setbox0\hbox{${\mathaccent"0362{#1}}^H$}%
  \setbox2\hbox{${\mathaccent"0362{\kern0pt#1}}^H$}%
  \ifdim\ht0=\ht2 #3\else #2\fi
  }
\newcommand*\rel@kern[1]{\kern#1\dimexpr\macc@kerna}
\newcommand*\widebar[1]{\@ifnextchar^{{\wide@bar{#1}{0}}}{\wide@bar{#1}{1}}}
\newcommand*\wide@bar[2]{\if@single{#1}{\wide@bar@{#1}{#2}{1}}{\wide@bar@{#1}{#2}{2}}}
\newcommand*\wide@bar@[3]{%
  \begingroup
  \def\mathaccent##1##2{%
    \let\mathaccent\save@mathaccent
    \if#32 \let\macc@nucleus\first@char \fi
    \setbox\z@\hbox{$\macc@style{\macc@nucleus}_{}$}%
    \setbox\tw@\hbox{$\macc@style{\macc@nucleus}{}_{}$}%
    \dimen@\wd\tw@
    \advance\dimen@-\wd\z@
    \divide\dimen@ 3
    \@tempdima\wd\tw@
    \advance\@tempdima-\scriptspace
    \divide\@tempdima 10
    \advance\dimen@-\@tempdima
    \ifdim\dimen@>\z@ \dimen@0pt\fi
    \rel@kern{0.6}\kern-\dimen@
    \if#31
      \overline{\rel@kern{-0.6}\kern\dimen@\macc@nucleus\rel@kern{0.4}\kern\dimen@}%
      \advance\dimen@0.4\dimexpr\macc@kerna
      \let\final@kern#2%
      \ifdim\dimen@<\z@ \let\final@kern1\fi
      \if\final@kern1 \kern-\dimen@\fi
    \else
      \overline{\rel@kern{-0.6}\kern\dimen@#1}%
    \fi
  }%
  \macc@depth\@ne
  \let\math@bgroup\@empty \let\math@egroup\macc@set@skewchar
  \mathsurround\z@ \frozen@everymath{\mathgroup\macc@group\relax}%
  \macc@set@skewchar\relax
  \let\mathaccentV\macc@nested@a
  \if#31
    \macc@nested@a\relax111{#1}%
  \else
    \def\gobble@till@marker##1\endmarker{}%
    \futurelet\first@char\gobble@till@marker#1\endmarker
    \ifcat\noexpand\first@char A\else
      \def\first@char{}%
    \fi
    \macc@nested@a\relax111{\first@char}%
  \fi
  \endgroup
}
\makeatother

\def\BibTeX{{\rm B\kern-.05em{\sc i\kern-.025em b}\kern-.08em
    T\kern-.1667em\lower.7ex\hbox{E}\kern-.125emX}}
\begin{document}

\title{Foreground-Background \\Ambient Sound Scene Separation\\
\thanks{This work was made with the support of the French National Research Agency, in the framework of the project LEAUDS ``Learning to understand audio scenes'' (ANR-18-CE23-0020). Experiments presented in this paper were carried out using the Grid'5000 testbed, supported by a scientific interest group hosted by Inria and including CNRS, RENATER and several Universities as well as other organizations (see \url{https://www.grid5000.fr}).}
}

\author{\IEEEauthorblockN{Michel Olvera\IEEEauthorrefmark{1}, Emmanuel Vincent\IEEEauthorrefmark{1}, Romain Serizel\IEEEauthorrefmark{1} and Gilles Gasso\IEEEauthorrefmark{2}}
\IEEEauthorblockA{\IEEEauthorrefmark{1}Universit\'e de Lorraine, CNRS, Inria, Loria, F-54000 Nancy, France}
\IEEEauthorblockA{\IEEEauthorrefmark{2}LITIS EA 4108, Universit\'e \& INSA Rouen Normandie, 76800 Saint-\'Etienne du Rouvray, France\\
Emails: \{michel.olvera, emmanuel.vincent\}@inria.fr, romain.serizel@loria.fr, gilles.gasso@insa-rouen.fr}}

\maketitle

\begin{abstract}
Ambient sound scenes typically comprise multiple short events occurring on top of a somewhat stationary background. We consider the task of separating these events from the background, which we call foreground-background ambient sound scene separation. We propose a deep learning-based separation framework with a suitable feature normalization scheme and an optional auxiliary network capturing the background statistics, and we investigate its ability to handle the great variety of sound classes encountered in ambient sound scenes, which have often not been seen in training. To do so, we create single-channel foreground-background mixtures using isolated sounds from the DESED and Audioset datasets, and we conduct extensive experiments with mixtures of seen or unseen sound classes at various signal-to-noise ratios. Our experimental findings demonstrate the generalization ability of the proposed approach.

\end{abstract}

\begin{IEEEkeywords}
Audio source separation, ambient sound scenes, generalization ability, deep learning.
\end{IEEEkeywords}

\section{Introduction}
\label{sec:Introduction}

Source separation aims at separating the constituent sources of a given recording \cite{vincent2018audio}. Current developments have benefited from
deep learning-based approaches, and outstanding results have been achieved for speech enhancement or separation \cite{weninger2015speech,yu2017permutation}. These methods have been rapidly extended to non-speech separation tasks such as music separation \cite{nugraha2016multichannel}.

It was not until recently that some works explored the separation of a much wider variety of sounds \cite{kavalerov2019universal, tzinis2019improving}, showing that deep neural networks can learn to separate short-duration mixtures of arbitrary sound classes. While this represents a big leap forward, real-life ambient sound analysis systems do not operate on short-duration mixtures of arbitrary sounds but on continuous recordings involving multiple short audio events in the presence of a somewhat stationary background. Potentially, audio surveillance systems could benefit from increased performance if, prior to classification and detection of audio events, short duration sounds of interest are isolated in a preprocessing step. Analogously, hearing aid devices could improve sound selectivity if audio events of interest are extracted or enhanced in the first place. In order to move toward this realistic setting, we study the separation of all short events (considered as a whole) from a stationary background, a task referred to in this paper as foreground-background ambient sound scene separation.

When separating a target source from another one belonging to a distinct class, such as in speech enhancement where the speech signal is to be separated from disruptive noise, a neural network is given the mixture signal or its time-frequency representation as input and is trained to estimate the target signal or a time-frequency mask \cite{kounovsky2017single, zhao2016dnn, doire2016single}. When separating a target source from another one belonging to the same class, such as in speaker extraction \cite{delcroix2018single} where a given target speaker is to be separated from interfering speakers and noise, this approach is not applicable anymore. It is customary in such a setting to assume that an isolated excerpt of the target source is available and to use an auxiliary network to summarize it into a vector which is provided to the main separation network \cite{vesely2016sequence}.

In this paper, we investigate whether a deep neural network can differentiate the rapidly varying spectro-temporal features of short audio events against the more slowly varying features of background sounds encountered in real-life environments, even when the foreground or background sound classes have not been seen in training. To do so, we rely on a deep neural network to estimate the soft time-frequency mask associated with short events, and incorporate an optional auxiliary network to inform it about the background statistics.


We provide three contributions. First, while auxiliary networks have been designed for speaker extraction \cite{delcroix2018single} where all sources belong to the same class (speech), we extend their application to situations when the sources in a mixture belong to distinct classes (here, ambient sounds). Secondly, we show that per-channel energy normalization (PCEN) of the mixture time-frequency representation \cite{wang2017trainable} improves the separation performance. Finally, we demonstrate the ability of the proposed approach to generalize to unseen foreground and/or background sounds. 

The remainder of this article is organized as follows. Section \ref{sec:Formulation} formulates the separation problem.  We present the proposed framework and the experimental setup in Sections \ref{sec:Framework} and   \ref{sec:ExperimentalSetup}. The achieved results and discussion are provided in Section \ref{sec:Results}. Finally, Section \ref{sec:Conclusions} concludes the paper.


\section{Problem formulation}
\label{sec:Formulation}
We define the problem of foreground-background separation in single-channel audio recordings as the task of recovering the foreground component, usually composed of sounds with rapidly varying spectral characteristics, in the presence of a slowly varying background component. The input mixture $x(t)$ is modeled as
\begin{equation}
  x(t) = f(t) + b(t)
\end{equation}
where
\begin{equation}
f(t)=\sum_{i=1}^{I}f_i(t) 
\end{equation}
with $I$ the number of foreground events and $\{f_i(t)\}_{i=1..I}$ the individual foreground events in the presence of a background sound $b(t)$ that is assumed to be stationary. Given the mixture $x(t)$, the goal is to estimate $f(t)$ and $b(t)$.

In this work, in order to draw clear conclusions, we focus on the recovery of a single foreground event overlapped with a single background sound, i.e., $I=1$. The analysis of more complex ambient mixtures is deferred to future work.

Separation is performed in the short-time Fourier transform (STFT) domain. The STFT coefficients $X(n,f)$ of the mixture $x(t)$ in time frame $n$ and frequency bin $f$ satisfy
\begin{equation}
    X(n,f) = F(n,f) + B(n,f)
\end{equation}
where $F(n,f)$ and $B(n,f)$ are the STFT coefficients of $f(t)$ and $b(t)$, respectively. 
We will use the notation $\mathbf{X}$, $\mathbf{F}$, $\mathbf{B}$ for the $N_x\times F$ matrices comprising all complex-valued coefficients $X(n,f)$, $F(n,f)$ and $B(n,f)$, with $N_x$ the number of time frames and $F$ the number of frequency bins. The STFT-domain components $\widehat{\mathbf{F}}$ and $\widehat{\mathbf{B}}$ estimated by the separation method are transformed into time-domain signals $\widehat{\mathbf{f}}(t)$ and $\widehat{\mathbf{b}}(t)$ by computing the inverse STFT.

\section{Separation Framework}
\label{sec:Framework}

\begin{figure}[htbp]
\centerline{\includegraphics[width=\columnwidth]{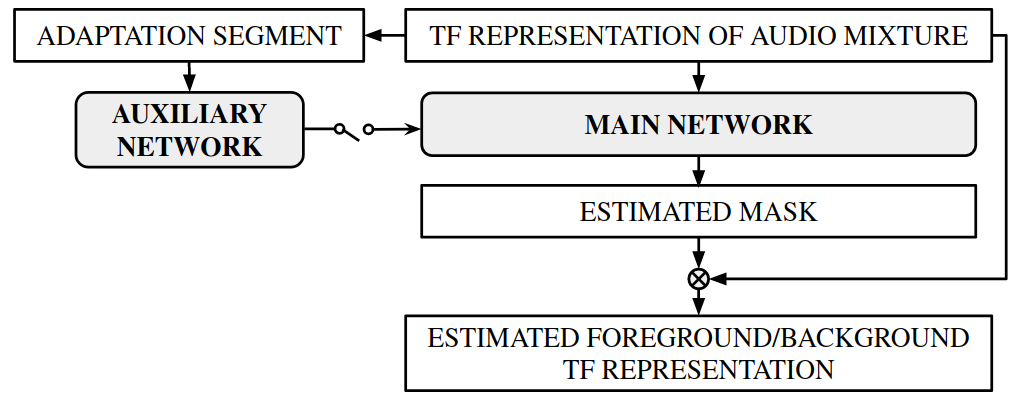}}
\caption{Block diagram of the proposed separation scheme.}
\label{fig:SeparationScheme}
\end{figure}

The proposed foreground-background separation framework is depicted in Fig.\ \ref{fig:SeparationScheme}. Inspired by the approach taken by \textit{SpeakerBeam} \cite{delcroix2018single} for single-channel target speaker extraction, it relies on a main deep neural network and an optional auxiliary network to locate background and foreground components in the time-frequency plane.

Classically, the two networks operate on the nonlinear Mel frequency scale \cite{wang2018supervised}. In the following, we denote by $F'$ the number of Mel bands and $|\mathbf{X}|^\text{Mel}$ the mixture Mel spectrogram which is obtained by multiplying the magnitude STFT $|\mathbf{X}|$ by the $F'\times F$ Mel filterbank matrix.

When the auxiliary network is \textit{inactive}, the main network takes the mixture log-Mel spectrogram $\log |\mathbf{X}|^\text{Mel}$ as input and outputs a time-frequency mask $\mathbf{M}^\text{Mel}$, i.e., an $N_x\times F'$ matrix with real-valued entries in $[0,1]$ that quantify the proportion of foreground sound in each time-frequency bin. This matrix is then projected back to the STFT-domain to obtain a $N_x\times F$ mask $\mathbf{M}$\footnote{We adopted this approach since directly outputting an STFT-domain mask did not make a significant difference.}. Using the mask, the estimated STFT magnitudes of the foreground and background components are obtained as
\begin{equation} \label{eq:ComponentEstimates}
    |\widehat{\mathbf{F}}| = \mathbf{M}\odot |\mathbf{X}| \quad\text{and}\quad |\widehat{\mathbf{B}}| = (\mathbf{1}-\mathbf{M})\odot |\mathbf{X}|,
\end{equation}
where $\odot$ denotes an element-wise multiplication and $\mathbf{1}$ is a matrix of ones. 

When the auxiliary network is \textit{active}, prior information about the background sound is assumed to be available in the form of an adaptation segment $a(t)$. Depending on the allowed latency, this could be a part of the mixture to be processed and/or a preceding time interval which has been classified with high confidence as background-only, i.e., without any overlapping foreground event. The auxiliary network compresses the $N_a \times F'$ log-Mel spectrogram of the adaptation segment $\log |\mathbf{A}|^\text{Mel}$, with $N_a$ the corresponding number of frames, into a fixed-size vector $\boldsymbol{\lambda}$ which is used together with the mixture log-Mel spectrogram $\log |\mathbf{X}|^\text{Mel}$ by the main network to output a time-frequency mask $\mathbf{M}^\text{Mel}$. This mask is converted to the STFT domain and used to estimate the STFT magnitudes of the foreground and background components via \eqref{eq:ComponentEstimates}. Such estimates are combined with the mixture phase to obtain the time-domain signals $\widehat{\mathbf{f}}(t)$ and $\widehat{\mathbf{b}}(t)$ by means of inverse STFT.


\section{Experimental setup}
\label{sec:ExperimentalSetup}

\subsection{Dataset}
In order to assess the generalization ability of the proposed framework in controlled conditions, we generated simulated foreground-background mixtures by randomly selecting isolated sounds from the development and evaluation sound banks of the Domestic Environment Sound Event Detection (DESED) dataset \cite{turpault2019sound} and Audioset \cite{gemmeke2017audio}, cutting them to 2~s length, and mixing them at various signal-to-noise ratios (SNRs) randomly chosen between -3 and 3~dB. Sounds shorter than 2~s are repeated as many times as needed.

The data are split
in three subsets for training, validation and evaluation. The corresponding sound classes are listed in Table \ref{tab:dataset_classes}. We considered a total of 25 sound classes: 10 classes of foreground events and 15 classes of background sounds. The isolated signals used to generate each subset are disjoint. 

\begin{table}[htbp]
\centering
\caption{Sound classes considered for the creation of the dataset.}\label{tab:dataset_classes}
\begin{tabularx}{\columnwidth}{|c|X|X|}
\hline
\textbf{Set} & \textbf{Foreground} & \textbf{Background} \\ 
\hline
\multirow{2}{*}{\textbf{\textit{Training}}} & \multirow{6}{=}{DESED: \texttt{\scriptsize dog, speech, cat, dishes, alarm-bell-ringing}} & \multirow{9}{=}{DESED: \texttt{\scriptsize vacuum cleaner, blender, frying, running water, electric shaver-toothbrush}, Audioset: \texttt{\scriptsize bathub, mechanical fan, microwave oven, hair dryer, drill}}  \\[1.125em]
\cline{1-1}
\multirow{2}{*}{\textbf{\textit{Validation}}} &  & \\ [1.125em]
\cline{1-1}
\multirow{2}{*}{\textbf{\textit{Evaluation C1}}} &  &  \\[1.125em] 
\cline{1-2}
\textbf{\textit{Evaluation C3}} & Audioset: \texttt{\scriptsize door, slam, squeak, coins, chopping food} & \\[1.125em] 
\hline
\textbf{\textit{Evaluation C2}} & DESED: \texttt{\scriptsize dog, speech, cat, dishes, alarm-bell-ringing} & \multirow{2}{=}{Audioset: \texttt{\scriptsize pink noise, white noise, noise, waterfall, vibration}} \\ 
\cline{1-2}
\textbf{\textit{Evaluation C4}} & Audioset: \texttt{\scriptsize door, slam, squeak, coins, chopping food} & \\ 
\hline
\end{tabularx}
\end{table}
For the training and validation sets, only 5 foreground classes from DESED and 10 background classes from Audioset and DESED are used. These \emph{seen} classes relate to domestic environments. The training data are augmented using Scaper \cite{salamon2017scaper}.

The evaluation set involves mixtures of these 15 seen classes and/or the remaining 10 \emph{unseen} classes, including 5 foreground classes and 5 background classes from Audioset. The unseen background classes are unrelated to domestic environments. The evaluation set comprises the following four subsets:
\begin{itemize}
    \item[\textit{C1}:] mixtures of seen foreground and background classes,
    \item[\textit{C2}:] mixtures of seen foreground classes and unseen background classes,
    \item[\textit{C3}:] mixtures of unseen foreground classes and seen background classes, 
    \item[\textit{C4}:] mixtures of fully unseen foreground and background classes.
\end{itemize}


Overall, the training set consists of $15,000$ mixtures (8.3 hours) generated from 604 isolated foreground events and 786 background events, along with $6,000$ mixtures for the validation set (3.3 hours) generated from 131 foreground events and 175 background events, and $1,000$ mixtures for each evaluation subset (0.5 hours) generated from 194 and 326 isolated foreground and background events, respectively. All data are single-channel signals sampled at 44.1 kHz.

\subsection{Input features and per-channel energy normalization}
We compute the STFT with a window size of 2048 samples (46 ms) and a hop size of 512 samples (12 ms), leading to an overlap of 75\% across frames. We then compute the log-Mel spectrogram using a Mel-filterbank of $F'=128$ filters.

In our first experiments, we found that the spectro-temporal structure of the foreground in the mixture log-Mel spectrogram is often less salient at higher frequencies, hence more poorly estimated. This 
led us to 
explore per-channel energy normalization (PCEN) \cite{lostanlen2018per}
as a way to enhance the foreground over the background. 
Instead of logarithmic compression, PCEN uses a simple feed-forward automatic gain control to dynamically stabilize signal levels. It is defined as
\begin{equation*}
    \text{PCEN}(n,f') = \left( \frac{|X|^\text{Mel}(n,f')}{(\epsilon + \widebar{|X|^\text{Mel}}(n,f'))^{\alpha}} + \delta \right)^{r} - \delta^{r},
\end{equation*}
where 
$f'$ denotes the Mel band index and $\widebar{|X|^\text{Mel}}(n,f')$ is a smoothed version of $|X|^\text{Mel}(n,f')$, which is computed using a first-order infinite impulse response (IIR) filter as
\begin{equation*}
    \widebar{|X|^\text{Mel}}(n,f') = (1 - s)\times \widebar{|X|^\text{Mel}}(n-1,f') + s\times |X|^\text{Mel}(n,f'),
\end{equation*}
with $s$ the smoothing constant.
This normalization scheme preserves frequency patterns while reducing stationary background sounds, thus making it a suitable acoustic front-end for the task at hand. We adopted the default PCEN parameters defined in \cite{lostanlen2018per}, i.e., $s=0.025$, $\epsilon=10^{-6}, \alpha=0.98, \delta=2$, and $r=0.5$, which are suited for indoor applications.

\subsection{Main network}
The neural network architecture is similar to what have been used in \cite{delcroix2018single} and \cite{hershey2016deep}. The main neural network comprises a stack of 3 recurrent layers with bidirectional long short-term (BLSTM) memory cells, each followed by a dense layer with an hyperbolic tangent activation function. Each BLSTM layer has 300 units in each direction and 0.2 dropout is applied on the output, which is then projected by the dense layer to 256 dimensions. The last layer is a dense layer with an output dimension of 128 that matches the number of frequency bands, and a sigmoid nonlinearity that ensures the output values are between 0 and 1. The network is trained on 170-frame segments using the Adam optimizer with learning rate 0.0001 for a total of 250 epochs.

\subsection{Auxiliary network}
The auxiliary network is a sequence-summarizing network \cite{vesely2016sequence}. It compresses the adaptation segment into a fixed-sized vector
$ \boldsymbol{\lambda} = \frac{1}{N_{a}}\sum_{n=1}^{N_a}{g(\log|\mathbf{a}|^\text{Mel}(n))}$,
where $|\mathbf{a}|^\text{Mel}(n)$ denotes the $n$-th frame of $|\mathbf{A}|^\text{Mel}$, and $g(\cdot)$ is the nonlinear transformation carried out by the network over each frame. This transformation comprises two dense layers with 128 neurons each and rectified linear unit (ReLU) activation functions.

The network outputs are averaged across all $N_a$ frames to produce the adaptation vector $\boldsymbol{\lambda}$. The entries of that vector are then used as scaling factors for the outputs of the first BLSTM layer of the main network. The auxiliary network is trained jointly with the main network. 

\subsection{Training objective}
Whether the auxiliary network is active or not, the model is trained by minimizing the squared Frobenius norm between the Mel spectrogram of the ground truth foreground event and the Mel spectrogram of the mixture multiplied element-wise by the estimated mask:
\begin{equation}
    \min \|\mathbf{M}^\text{Mel} \odot |\mathbf{X}|^\text{Mel} - |\mathbf{F}|^\text{Mel}\|^{2}_F.
\end{equation}

\begin{figure*}[htbp]
\centerline{\includegraphics[width=1\textwidth]{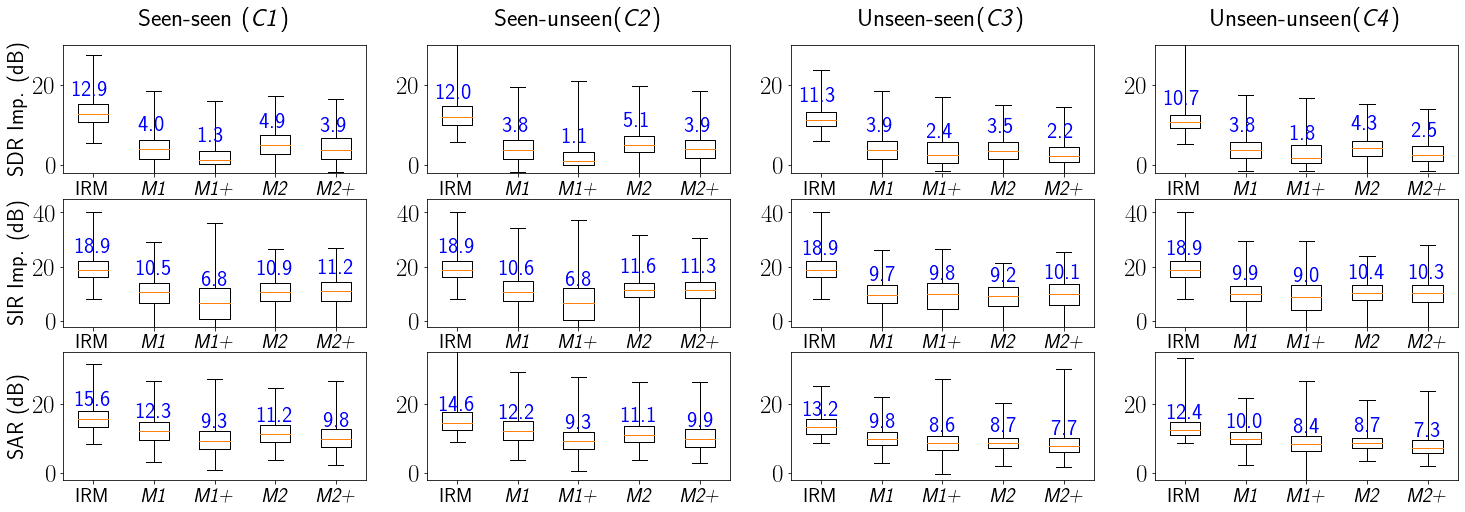}}
\caption{SDR and SIR improvements and SAR (dB) achieved by the four model configurations and the IRM on the four subsets of the evaluation set.}
\label{fig:SeparationResults}
\end{figure*}

\subsection{Model configurations}
We built four models to investigate the separation task, which are briefly described below. The first model, referred to as \textit{M1}, is trained with log-Mel spectrograms as inputs
and does not use the auxiliary network in the mask estimation process. The second model, \textit{M1+}, is similar to \textit{M1} and makes use of the auxiliary network.
Models \textit{M2} and \textit{M2+} are defined similarly but use PCEN spectrograms as inputs instead of log-Mel spectrograms.

For comparison, we also computed the ideal ratio mask (IRM) $\mathbf{M}^\text{IRM} = \mathbf{F}^\text{Mel}/ (\mathbf{F}^\text{Mel} + \mathbf{B}^\text{Mel})$,
where $\mathbf{F}^\text{Mel}$ and $\mathbf{B}^\text{Mel}$ are the Mel spectrograms of the ground truth foreground and background signals. We used Mel spectrograms as the time-frequency representation for the computation of the IRM such that the corresponding results provide an upper bound for the performance of the foreground-background separation task.

\section{Results}
\label{sec:Results}
As objective criteria to evaluate the quality of the separation, we used the signal-to-distortion ratio (SDR), source-to-interference ratio (SIR) and signal-to-artifacts (SAR) metrics defined by \cite{vincent2006performance}. Fig. \ref{fig:SeparationResults} reports the SDR and SIR improvements  and the SAR achieved by the four model configurations and the IRM on the four subsets of the evaluation set.


\subsection{Impact of feature normalization and the auxiliary network}
Model \textit{M2} achieved a higher median SDR improvement than other models on subsets \textit{C1},  \textit{C2} and \textit{C4}. In particular, compared to model \textit{M1}, it increased the median SDR improvement by 0.9, 1.3 and 0.5 dB for \textit{C1}, \textit{C2} and \textit{C4}, respectively. This indicates that PCEN improves the separation performance in terms of SDR and SIR. However the performance of model \textit{M2} decreases for mixtures with unseen foreground classes, while that of model \textit{M1} shows less variability. Regarding models using the auxiliary network, \textit{M1+} and \textit{M2+} perform worse than their counterparts \textit{M1} and \textit{M2} in terms of SDR. In fact, model \textit{M1+} is the worst among all models. Model \textit{M2+} showed similar performance than \textit{M2} in terms of SIR, but its lower SAR scores impacted negatively its SDR performance. However, \textit{M2+} can relate to  model \textit{M1} on subsets with seen foreground. Since the sources to be separated are of substantially different classes, the use of an auxiliary network may not be necessary. This stands in contrast with the speaker extraction task in which the signals to be separated are from the same class, and an auxiliary network is beneficial.

\subsection{Robustness to unseen events}
Robustness to unseen foreground-background classes can be analyzed by comparing the four subplots in each row of Fig. \ref{fig:SeparationResults} with each other. The SDR improvement achieved by model \textit{M1} is persistent over the four evaluation subsets, showing good generalization for any combination of either seen or unseen foreground-background sound classes in terms of SDR. Model \textit{M2} improves SDR scores, but shows less robustness to unseen sound classes, nevertheless its performance on subsets \textit{C3} and \textit{C4}, can be compared to that of model \textit{M1}.

\subsection{Signal distortion}
\begin{figure*}[htbp]
\centerline{\includegraphics[width=1\textwidth]{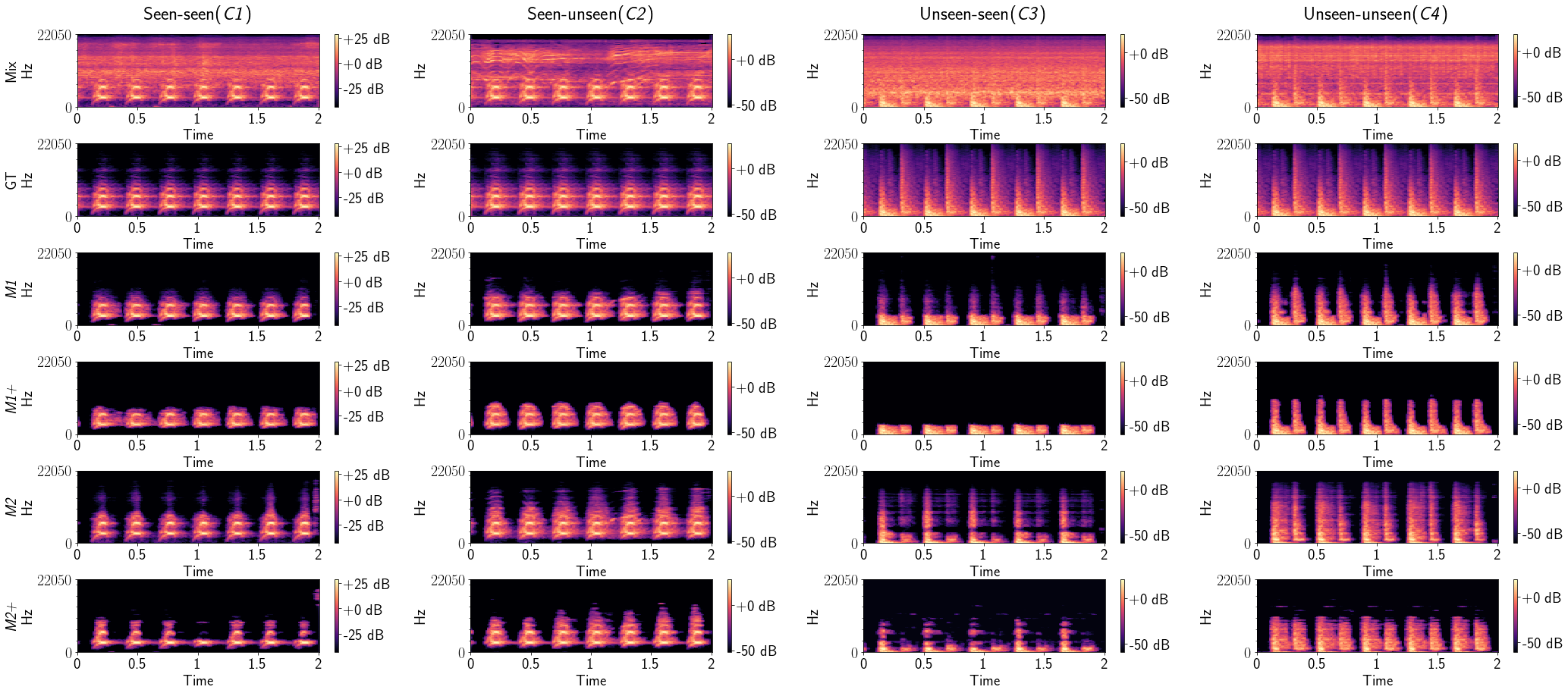}}
\caption{Examples of foreground estimates obtained using the four model configurations on the four subsets of the evaluation set. GT stands for ground truth.}
\label{fig:SeparationExamples}
\end{figure*}

Fig. \ref{fig:SeparationExamples} illustrates the effect of the auxiliary network from a qualitative perspective. Adding the adaptation segment to the main network causes the background sound to be more strongly reduced, leading to strong distortion of the foreground (\textit{M1+}). PCEN improves the results (\textit{M2+}). Yet, the separation results for models \textit{M1} and \textit{M2} confirm that better separation quality is achieved when the auxiliary network is inactive. The reader may visit the accompanying website\footnote{\url{http://molveraz.com/ambient-sound-scene-separation/}} for audio examples. 

\section{Conclusion}
\label{sec:Conclusions}
We presented the foreground-background ambient sound separation task, in which short duration events occur on top of a background sound. This task is closer to the conditions faced by real-life ambient sound analysis systems than the mixtures of arbitrary sounds considered in some previous studies. We performed a series of experiments to assess the performance of a deep learning-based 
model to discriminate the rapidly varying spectro-temporal features of foreground events against the slowly varying features of background sounds. We considered a scheme comprising a main mask estimation network and analyzed the effects of feeding an auxiliary network with an adaptation segment. Also, we explored the impact of feature normalization. Under the proposed experimental protocol, we found that the use of adaptation segments to inform the network is detrimental to the separation process while the use of PCEN is beneficial. The similar improvements achieved by the proposed model over mixtures of seen and unseen sound classes show its generalization capabilities in terms of SDR.

\bibliographystyle{IEEEtran}
\bibliography{bibliography.bib}

\begin{thebibliography}{10}
\providecommand{\url}[1]{#1}
\csname url@samestyle\endcsname
\providecommand{\newblock}{\relax}
\providecommand{\bibinfo}[2]{#2}
\providecommand{\BIBentrySTDinterwordspacing}{\spaceskip=0pt\relax}
\providecommand{\BIBentryALTinterwordstretchfactor}{4}
\providecommand{\BIBentryALTinterwordspacing}{\spaceskip=\fontdimen2\font plus
\BIBentryALTinterwordstretchfactor\fontdimen3\font minus
  \fontdimen4\font\relax}
\providecommand{\BIBforeignlanguage}[2]{{%
\expandafter\ifx\csname l@#1\endcsname\relax
\typeout{** WARNING: IEEEtran.bst: No hyphenation pattern has been}%
\typeout{** loaded for the language `#1'. Using the pattern for}%
\typeout{** the default language instead.}%
\else
\language=\csname l@#1\endcsname
\fi
#2}}
\providecommand{\BIBdecl}{\relax}
\BIBdecl

\bibitem{vincent2018audio}
E.~Vincent, T.~Virtanen, and S.~Gannot, \emph{Audio Source Separation and
  Speech Enhancement}.\hskip 1em plus 0.5em minus 0.4em\relax John Wiley \&
  Sons, 2018.

\bibitem{weninger2015speech}
F.~Weninger, H.~Erdogan, S.~Watanabe, E.~Vincent, J.~Le~Roux, J.~R. Hershey,
  and B.~Schuller, ``Speech enhancement with {LSTM} recurrent neural networks
  and its application to noise-robust {ASR},'' in \emph{International
  Conference on Latent Variable Analysis and Signal Separation}, 2015, pp.
  91--99.

\bibitem{yu2017permutation}
D.~Yu, M.~Kolb{\ae}k, Z.-H. Tan, and J.~Jensen, ``Permutation invariant
  training of deep models for speaker-independent multi-talker speech
  separation,'' in \emph{2017 IEEE International Conference on Acoustics,
  Speech and Signal Processing (ICASSP)}, 2017, pp. 241--245.

\bibitem{nugraha2016multichannel}
A.~A. Nugraha, A.~Liutkus, and E.~Vincent, ``Multichannel music separation with
  deep neural networks,'' in \emph{24th European Signal Processing Conference
  (EUSIPCO)}, 2016, pp. 1748--1752.

\bibitem{kavalerov2019universal}
I.~Kavalerov, S.~Wisdom, H.~Erdogan, B.~Patton, K.~Wilson, J.~Le~Roux, and
  J.~R. Hershey, ``Universal sound separation,'' in \emph{2019 IEEE Workshop on
  Applications of Signal Processing to Audio and Acoustics (WASPAA)}, 2019, pp.
  175--179.

\bibitem{tzinis2019improving}
E.~Tzinis, S.~Wisdom, J.~R. Hershey, A.~Jansen, and D.~P.~W. Ellis, ``Improving
  universal sound separation using sound classification,'' \emph{arXiv preprint
  arXiv:1911.07951}, 2019.

\bibitem{kounovsky2017single}
T.~Kounovsky and J.~Malek, ``Single channel speech enhancement using
  convolutional neural network,'' in \emph{2017 IEEE International Workshop of
  Electronics, Control, Measurement, Signals and their Application to
  Mechatronics (ECMSM)}, 2017, pp. 1--5.

\bibitem{zhao2016dnn}
Y.~Zhao, D.~Wang, I.~Merks, and T.~Zhang, ``{DNN}-based enhancement of noisy
  and reverberant speech,'' in \emph{2016 IEEE International Conference on
  Acoustics, Speech and Signal Processing (ICASSP)}, 2016, pp. 6525--6529.

\bibitem{doire2016single}
C.~S.~J. Doire, M.~Brookes, P.~A. Naylor, C.~M. Hicks, D.~Betts, M.~A. Dmour,
  and S.~H. Jensen, ``Single-channel online enhancement of speech corrupted by
  reverberation and noise,'' \emph{IEEE/ACM Transactions on Audio, Speech, and
  Language Processing}, vol.~25, no.~3, pp. 572--587, 2016.

\bibitem{delcroix2018single}
M.~Delcroix, K.~\v{Z}mol\'ikov\'a, K.~Kinoshita, A.~Ogawa, and T.~Nakatani,
  ``Single channel target speaker extraction and recognition with speaker
  beam,'' in \emph{2018 IEEE International Conference on Acoustics, Speech and
  Signal Processing (ICASSP)}, 2018, pp. 5554--5558.

\bibitem{vesely2016sequence}
K.~Vesel{\'y}, S.~Watanabe, K.~{\v{Z}}mol{\'\i}kov{\'a}, M.~Karafi{\'a}t,
  L.~Burget, and J.~H. {\v{C}}ernock{\'y}, ``Sequence summarizing neural
  network for speaker adaptation,'' in \emph{2016 IEEE International Conference
  on Acoustics, Speech and Signal Processing (ICASSP)}, 2016, pp. 5315--5319.

\bibitem{wang2017trainable}
Y.~Wang, P.~Getreuer, T.~Hughes, R.~F. Lyon, and R.~A. Saurous, ``Trainable
  frontend for robust and far-field keyword spotting,'' in \emph{2017 IEEE
  International Conference on Acoustics, Speech and Signal Processing
  (ICASSP)}, 2017, pp. 5670--5674.

\bibitem{wang2018supervised}
D.~Wang and J.~Chen, ``Supervised speech separation based on deep learning: An
  overview,'' \emph{IEEE/ACM Transactions on Audio, Speech, and Language
  Processing}, vol.~26, no.~10, pp. 1702--1726, 2018.

\bibitem{turpault2019sound}
N.~Turpault, R.~Serizel, A.~Parag~Shah, and J.~Salamon, ``{Sound event
  detection in domestic environments with weakly labeled data and soundscape
  synthesis},'' in \emph{Workshop on Detection and Classification of Acoustic
  Scenes and Events (DCASE)}, 2019.

\bibitem{gemmeke2017audio}
J.~F. Gemmeke, D.~P.~W. Ellis, D.~Freedman, A.~Jansen, W.~Lawrence, R.~C.
  Moore, M.~Plakal, and M.~Ritter, ``Audio set: An ontology and human-labeled
  dataset for audio events,'' in \emph{2017 IEEE International Conference on
  Acoustics, Speech and Signal Processing (ICASSP)}, 2017, pp. 776--780.

\bibitem{salamon2017scaper}
J.~Salamon, D.~MacConnell, M.~Cartwright, P.~Li, and J.~P. Bello, ``Scaper: A
  library for soundscape synthesis and augmentation,'' in \emph{2017 IEEE
  Workshop on Applications of Signal Processing to Audio and Acoustics
  (WASPAA)}, 2017, pp. 344--348.

\bibitem{lostanlen2018per}
V.~Lostanlen, J.~Salamon, M.~Cartwright, B.~McFee, A.~Farnsworth, S.~Kelling,
  and J.~P. Bello, ``Per-channel energy normalization: Why and how,''
  \emph{IEEE Signal Processing Letters}, vol.~26, no.~1, pp. 39--43, 2018.

\bibitem{hershey2016deep}
J.~R. Hershey, Z.~Chen, J.~Le~Roux, and S.~Watanabe, ``Deep clustering:
  Discriminative embeddings for segmentation and separation,'' in \emph{2016
  IEEE International Conference on Acoustics, Speech and Signal Processing
  (ICASSP)}.\hskip 1em plus 0.5em minus 0.4em\relax IEEE, 2016, pp. 31--35.

\bibitem{vincent2006performance}
E.~Vincent, R.~Gribonval, and C.~F{\'e}votte, ``Performance measurement in
  blind audio source separation,'' \emph{IEEE Transactions on Audio, Speech,
  and Language Processing}, vol.~14, no.~4, pp. 1462--1469, 2006.

\end{thebibliography}
\end{document}